# High electrical conducting deep-ultraviolet-transparent oxide semiconductor La-doped SrSnO$_3$ exceeding ~3000 S cm$^{-1}$

Mian Wei[1], Anup V. Sanchela[2], Bin Feng[3], Yuichi Ikuhara[3], Hai Jun Cho[1,2*], and Hiromichi Ohta[1,2*]

[1]*Graduate School of Information Science and Technology, Hokkaido University, N14W9, Kita, Sapporo 060−0814, Japan*

[2]*Research Institute for Electronic Science, Hokkaido University, N20W10, Kita, Sapporo 001−0020, Japan*

[3]*Institute of Engineering Innovation, The University of Tokyo, 2−11−16 Yayoi, Bunkyo, Tokyo 113−8656, Japan*

\*Authors to whom correspondence should be addressed:

Hai Jun Cho (joon@es.hokudai.ac.jp) and Hiromichi Ohta (hiromichi.ohta@es.hokudai.ac.jp)

**La-doped SrSnO$_3$ (LSSO) is known as one of deep-ultraviolet (DUV)-transparent conducting oxides with an energy bandgap of ~4.6 eV. Since LSSO can be grown heteroepitaxially on more wide bandgap substrates such as MgO ($E_g$ ~7.8 eV), LSSO is considered to be a good candidate as a DUV-transparent electrode. However, the electrical conductivity of LSSO films are below 1000 S cm$^{-1}$, most likely due to the low solubility of La ion in the LSSO lattice. Here we report that high electrically conducting (>3000 S cm$^{-1}$) LSSO thin films with an energy bandgap of ~4.6 eV can be fabricated by pulsed laser deposition on MgO substrate followed by a simple annealing in vacuum. From the X-ray diffraction and the scanning transmission electron microscopy analyses, we found that lateral grain growth occurred during the annealing, which improved the** **activation rate of La ion, leading to a significant improvement of carrier concentration ($3.26 \times 10^{20}$ cm$^{-3}$) and Hall mobility (55.8 cm$^2$ V$^{-1}$ s$^{-1}$). The present DUV-transparent oxide semiconductor would be useful as a transparent electrode for developing optoelectronic devices, which transmit and/or emit DUV-light.**

Deep-ultraviolet (DUV, 200−300 nm in wavelength) transparent oxide semiconductors[1] have attracted attention as transparent electrodes for DUV optoelectronic devices such as DUV light emitting diode and DUV detector for biological applications because most DNA show their absorption peaking around 260 nm in wavelength[2]. Although conventional transparent oxide semiconductors such as Al-doped ZnO and Sn-doped In$_2$O$_3$ (ITO) are opaque in DUV region (wavelength $\lambda$ < 300 nm) due to their small bandgaps ($E_g$ ~3.2 eV), DUV transparent oxide semiconductors are transparent in DUV region. Among several DUV transparent oxide semiconductors, La-doped SrSnO$_3$ ($E_g$ ~4.6 eV[3-6], LSSO hereafter) would be the most promising material because LSSO can be grown heteroepitaxially on the single crystalline substrates such as MgO[7], SrTiO$_3$[8], and KTaO$_3$[9]. Further, the electrical conductivity of LSSO films (~1000 S cm$^{-1}$)[10] is larger than well-known DUV transparent oxide semiconductors such as β-Ga$_2$O$_3$ ($E_g$ ~4.9 eV, ~1 S cm$^{-1}$)[11,12], α-Ga$_2$O$_3$ ($E_g$ ~5.3 eV, ~0.3 S cm$^{-1}$)[13], electron-doped calcium aluminate (C12A7:$e^-$, ~4 eV, ~800 S cm$^{-1}$)[14-16], and recently developed Al:Mg$_{0.43}$Zn$_{0.57}$O ($E_g$ ~4.2 eV, ~400 S cm$^{-1}$)[17]. However, the electrical conductivity of LSSO films is still lower than that of commercially available transparent conducting oxide ITO (~7000 S cm$^{-1}$).

The crystal structure of SrSnO$_3$ is regarded to pseudo-double-cubic perovskite ($a$ = 0.8068 nm)[18] though the real lattice is orthorhombic[19]. Several researchers fabricated the epitaxial films of LSSO by using vapor phase epitaxy method such as pulsed laser deposition (PLD)[8-10] and molecular beam epitaxy (MBE)[20]. However, due to difficulties in growing high-quality

LSSO films, their electron transport properties have not been extensively studied compared to the optical properties and electronic structures. Baba *et al.*[10] reported that the electrical conductivity of 5% La-doped LSSO film on lattice matched NdScO$_3$ substrate by PLD was ~1000 S cm$^{-1}$, which is the largest electrical conductivity ever reported. However, the carrier concentration was only $2.5 \times 10^{20}$ cm$^{-3}$, which is ~1/3 of La concentration [La], indicating low activation rate of La. We hypothesized that the electrical conductivity of LSSO films can be improved if the solubility of La ions in the LSSO lattice is enhanced.

In order to enhance the solubility of La ions in the LSSO films, we simply annealed the LSSO films at 790 °C in vacuum (<10$^{-2}$ Pa). Generally, sintering of n-type oxide semiconductor becomes faster when the point defect (oxygen vacancy) is formed. For example, Al-doped ZnO is generally sintered in reducing atmosphere. We expected that the point defect formation during the annealing in vacuum would induce grain growth and increase the solubility of La ion. After annealing, we indeed observed grain growth of PLD-grown LSSO films from ~10 nm to ~28 nm, and the electron transport properties of the annealed LSSO films were significantly improved. Here we report that high electrically conducting (>3000 S cm$^{-1}$) LSSO thin films were fabricated by pulsed laser deposition on MgO ($E_g$ ~7.8 eV[21]) substrate followed by the simple annealing in vacuum. The significant improvement of carrier concentration ($3.26 \times 10^{20}$ cm$^{-3}$) and Hall mobility (55.8 cm$^2$ V$^{-1}$ s$^{-1}$) occurred after the annealing. The present DUV-transparent oxide semiconductor would be useful as the transparent electrode to develop optoelectronic devices, which transmit and/or emit DUV-light.

La$_x$Sr$_{1-x}$SnO$_3$ ($x$ = 0.005, 0.02, 0.03 and 0.05) epitaxial films were fabricated on (001) MgO ($a$ = 0.4212 nm) single crystal substrates using PLD technique (KrF excimer laser, $\lambda$ = 248 nm, fluence ~2 J cm$^{-2}$ pulse$^{-1}$, repetition rate = 10 Hz). During the film growth, substrate temperature and oxygen pressure inside the chamber were kept at 790 °C and 20 Pa, respectively. Note that conducting LSSO film was not obtained when the oxygen pressure was lower than 20 Pa, indicating that oxygen vacancies do not play as the electron donor, similar to La-doped BaSnO$_3$[22]. After the film growth, we turned off the substrate heater immediately and cooled the sample down to room temperature. The crystalline phase, orientation, lattice parameters, and thickness of the films were analyzed by high resolution X-ray diffraction (Cu Kα$_1$, ATX-G, Rigaku Co.). Out-of-plane Bragg diffraction patterns and the rocking curves were measured at room temperature. X-ray reflection patterns were measured to evaluate the density and the thickness. Atomic force microscopy (AFM, Nanocute, Hitachi Hi-Tech Sci. Co.) was used to observe the surface microstructure of the films.

In order to clarify the change of the lateral grain size, we measured X-ray reciprocal space mapping (RSM) of the resultant LSSO film ($x = 0.03$, as grown) around 204 diffraction spot of LSSO [**Fig. 1(a)**]. Broad diffraction spot of 204 LSSO is seen peaking at ($q_x/2\pi$, $q_z/2\pi$) = (−4.97 nm$^{-1}$, 9.88 nm$^{-1}$) together with 204 diffraction spot of MgO at (−4.75 nm$^{-1}$, 9.50 nm$^{-1}$), indicating incoherent epitaxial growth occuured. The lateral grain size was ~12 nm, which was calculated by Scherrer's equation, $D =$ (Integral width)$^{-1}$, using the cross sectional diffraction pattern along $q_x/2\pi$ direction.

In order to increase the lateral grain size, we annealed the LSSO films at 790 °C in vacuum (<10$^{-2}$ Pa) for 30 min and cooled down to room temperature. After the vacuum annealing, the 204 diffraction spot intensity increased as shown in **Fig. 1(b)**, and the lateral grain size was dramatically increased up to 28 nm. Similar increase of the diffraction spot intensity was observed in all cases ($x = 0.005$, 0.02 and 0.05) as shown in **Supplementary Figs. S1(a)−S1(h)**. Using the 204 diffraction spots, we determined the average lattice parameters, $(a^2 \cdot c)^{1/3}$ of the LSSO films before and after annealing [**Fig. 1(c)**]. The annealed films showed more relaxation, which is likely attributed to the grain growth and the improvements in the solubility of La ion. Note that we optimized the annealing condition by annealing under several atmosphere including air, 1 Pa oxygen, and vacuum ($<10^{-2}$ Pa) (data not shown). Significant grain growth was observed when the LSSO film was annealed in vacuum. This is consistent with the fact that the accelerated dislocation movement occurs due to oxygen vacancy-assisted recovery in ionic crystals[23]. Therefore, we fixed the annealing atmosphere as vacuum. According to previous studies on stannate films, extra charge carriers from oxygen vacancies are compensated by the formation of charge accepting A-site vacancies.[24,25] Therefore, any carrier concentration changes that may result from the vacuum annealing are likely attributed to the activation of La dopants.

In order to further confirm the grain growth due to the vacuum annealing, the microstructures of the LSSO films were observed using low-angle annular dark-field scanning transmission electron microscopy (LAADF-STEM). While columnar structures were observed for both the as-deposited [**Fig. 2(a)**] and annealed LSSO films [**Fig. 2(b)**], the width of the columns for the annealed LSSO film was significantly greater. This indicates that the vacuum annealing indeed induces grain growth, which is in excellent agreement with the RSM results. It should be noted that incoherent interfaces with misfit dislocations was observed at the interface between the LSSO film and MgO substrate in both cases as shown in **Supplementary Figs. S2(a)** and **S2(b)**, indicating that the interface structures are not affected by the vacuum annealing.

Then, the electrical conductivity ($\sigma$), carrier concentration ($n$), and Hall mobility ($\mu_{Hall}$) of the films were measured using the conventional dc four-probe method with van der Pauw electrode geometry at room temperature. The thermopower ($S$) was acquired from the thermo-electromotive force ($\Delta V$) generated by a temperature difference ($\Delta T$) of ~4 K across the film using two Peltier devices. The temperatures at each end of the films were simultaneously measured with two thermocouples, and the $S$-values were calculated from the slope of the $\Delta T-\Delta V$ plots (correlation coefficient: > 0.9999).

**Figure 3** summarizes the electron transport properties of the resultant LSSO films at room temperature. Both carrier concentration and Hall mobility gradually increased with [La] until [La] = 3% and decreased at [La] = 5%, probably due to the low solubility limit of La in $SrSnO_3$ lattice. This is similar to $BaSnO_3$[26]. Significant enhancements in the electron transport properties were observed in the vacuum annealed LSSO films. Compared to the as grown LSSO films, the annealed LSSO films showed greater $\sigma$ [**Fig. 3(a)**], $n$ [**Fig. 3(b)**], and $\mu_{Hall}$ [**Fig. 3(d)**] at room temperature. The largest enhancement was observed in 5% La doped LSSO film. The absolute value of $S$ for LSSO films decreased after annealing [**Fig. 3(c)**]. This is consistent with the fact that carrier concentration increased after annealing. The activation rate of La ion [**Fig. 3(e)**], which was calculated as $n \cdot [La]^{-1}$ assuming $La^{3+}$ ion solely generate carrier electrons, increased after the vacuum annealing. Thus, the increased $\sigma$ and $\mu_{Hall}$ of annealed films are probably attributed to the improved efficiency of the carrier activation from the La dopant and the grain growth. Therefore, carrier electrons are more efficiently provided in the annealed films. The highest $\sigma$ and $\mu_{Hall}$ was observed at annealed 3%-La doped LSSO films, whose activation of La ion and lateral grain size showed largest value.

In this study, the experimental electron effective mass ($m^*$) value of the LSSO epitaxial films was also clarified by the relationship of thermopower ($S$) and carrier concentration ($n$) of resultant films at room temperature [**Fig. 4(a)**]. Since $n$ and $S$ value both are only sensitive to the density of states and Fermi level, we calculated the $m^*$ using the following equations.[27]

$$S = -\frac{k_B}{e}\left(\frac{(r+2)F_{r+1}(\xi)}{(r+1)F_r(\xi)} - \xi\right) \quad (1)$$

where $k_B$, $\xi$, $r$, and $F_r$ are the Boltzmann constant, reduced Fermi energy, scattering parameter of relaxation time, and Fermi integral, respectively. We assumed the r value of 0.5 (acoustic phonon scattering) because the temperature dependence of Hall mobility showed that phonon scattering is dominant at room temperature (data not shown). $F_r$ and $n$ are given by,

$$F_r(\xi) = \int_0^\infty \frac{x^r}{1+e^{x-\xi}} dx \quad (2)$$

$$n_- = 4\pi \left(\frac{2m^* k_B T}{h^2}\right)^{3/2} F_{1/2}(\xi) \quad (3)$$

where $h$ and $T$ are the Planck constant and absolute temperature, respectively. Most of the $S - n$ plots ($n > 10^{20}$ cm$^{-3}$) are located on the line obtained $m^*$ of 0.23 $m_0$, similar to that of $Ba_{0.5}Sr_{0.5}SnO_3$[28], though two plots ($n < 10^{20}$ cm$^{-3}$) are located lower side, probably due to the contribution of the tail state around the conduction band minimum.

**Figure 4(b)** summarizes the Hall mobility ($\mu_{Hall}$) of the LSSO epitaxial films as a function of carrier concentration ($n$) at room temperature. The values reported by Wang et al.[20] and Baba et al.[10] are also plotted for comparison. The $\mu_{Hall}$ increased with increasing $n$. The highest $\mu_{Hall}$ and electrical conductivity were observed from 3% La-doped LSSO film. The $\mu_{Hall}$ reached 55.8 cm$^2$ V$^{-1}$ s$^{-1}$ at a carrier concentration of $3.26 \times 10^{20}$ cm$^{-3}$, which is highest value ever reported in the LSSO film to date. Moreover, we achieved an electrical conductivity exceeding 3000 S cm$^{-1}$ in annealed 3% La-doped LSSO film, which is much higher than that of other DUV transparent conducting oxides such as electron doped β-Ga$_2$O$_3$ ($E_g$ ~4.9 eV, ~1 S cm$^{-1}$)[11,12], α-Ga$_2$O$_3$ ($E_g$ ~5.3 eV, ~0.3 S cm$^{-1}$)[13], C12A7:$e^-$, ~4 eV, ~800 S cm$^{-1}$)[14-16], and recently developed Al:Mg$_{0.43}$Zn$_{0.57}$O ($E_g$ ~4.2 eV, ~400 S cm$^{-1}$)[17].

Finally we measured the optical transmission and reflection spectra of the 3% La-doped LSSO film (thickness: 112 nm) grown of double side polished (001) MgO substrate followed by the vacuum annealing at 790 °C **(inset of Fig. 5)**. The optical transmission in the deep UV region was kept ~70 % at 300 nm in wavelength and exceeded 25 % at 260 nm in wavelength, which is the absorption peak wavelength of most DNAs.[2] Although $SrSnO_3$ has an indirect bandgap,[29] we estimated the bandgap by assuming a direct bandgap; $(\alpha \cdot h \cdot v)^2$−photon energy plot was used to estimate the bandgap of the LSSO film **(Fig. 5)**, where $\alpha$ is the absorption coefficient and $v$ is speed of light. The bandgap was ~4.6 eV, agreeing well with the reported values[3-5]. The $\alpha$ at 260 nm in wavelength was calculated to be $8.3 \times 10^4$ cm$^{-1}$; the penetration depth of 260 nm light is ~120 nm, which is the origin of low transmission (25%) of the 260 nm light through the LSSO film (112 nm). Thus, most of the 260 nm light can transmit through LSSO film when the thickness is thinner. From these results, we concluded that present LSSO film, which exhibited high electrical conductivty of ~3000 S cm$^{-1}$ and deep UV transparency, would be useful as the transparent electrode to develop optoelectronic devices, which transmit and/or emit DUV-light especially for biological applications.

In summary, we fabricated high electrically conducting (>3000 S cm$^{-1}$) LSSO thin films with an energy bandgap of ~4.6 eV by a pulsed laser deposition on MgO substrate followed by the simple annealing in vacuum. From the X-ray diffraction and the scanning transmission microscopy analyses, we found that lateral grain growth occurred during the annealing, which improves the activation rate of La ion, leading to the significant improvement of carrier concentration ($3.26 \times 10^{20}$ cm$^{-3}$) and Hall mobility (55.8 cm$^2$ V$^{-1}$ s$^{-1}$). The bandgap was ~4.6 eV. The optical transmission at 300 nm in wavelength was ~70 % and that at 260 nm in wavelength exceeded 25 %. The present DUV-transparent oxide semiconductor would be useful as the transparent electrode to develop optoelectronic devices which transmit and/or emit DUV-light especially for biological applications.

**Supplementary Material**

See supplementary material for the RSMs and the cross-sectional HAADF-STEM images of the LSSO films.

**Acknowledgements**

This research was supported by Grants-in-Aid for Scientific Research A (17H01314) and Innovative Areas (19H05791) from the JSPS. A part of this work was supported by Dynamic Alliance for Open Innovation Bridging Human, Environment, and Materials, and by the Network Joint Research Center for Materials and Devices. H.J.C. acknowledges the support from Nippon Sheet Glass Foundation for Materials Science and Engineering. H.O. acknowledges the support from the Asahi Glass Foundation and the Mitsubishi Foundation The support from China Scholarships Council (M. Wei, 201808050081) is also greatly appreciated.

**Contributions**

M.W. and H.J.C. performed the sample preparation and measurements. B.F. and Y.I. performed the STEM analyses. H.O. planned and supervised the project. All authors discussed the results and commented on the manuscript.

**Competing financial interests**

The authors declare no competing financial interests.

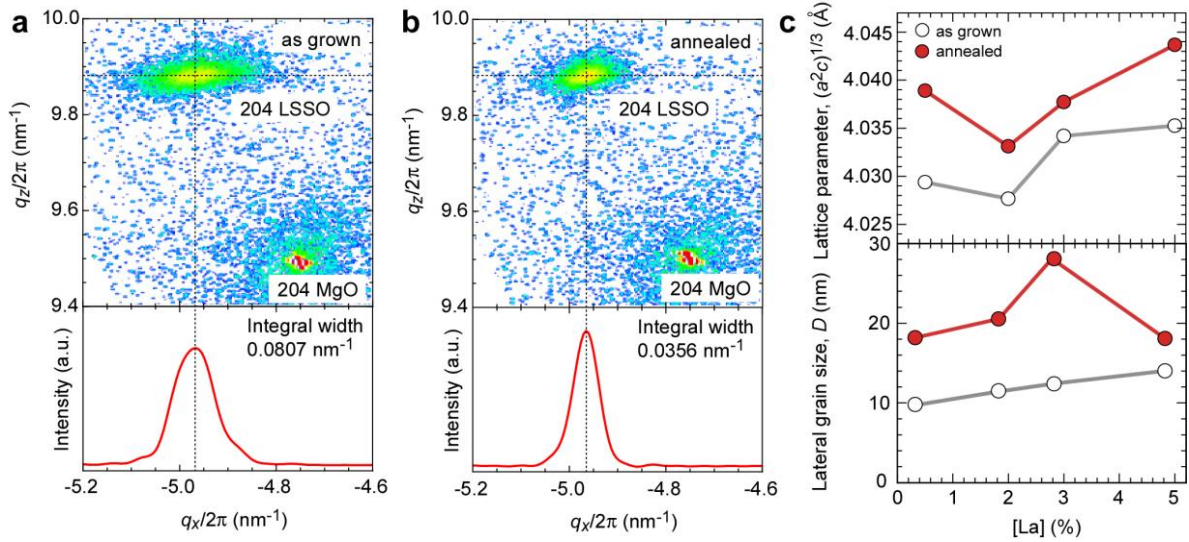

**FIG. 1** | **Lateral grain growth of the LSSO films by vacuum annealing.** (a, b) RSMs around 204 LSSO diffraction spot of (a) as grown and (b) annealed films. The cross sectional diffraction patterns along $q_x/2\pi$ direction are shown at the bottom of each RSM. (c) Changes in (upper) the average lattice parameter and (lower) lateral grain size of the as-deposited LSSO films (white) and annealed LSSO films (red) as a function of La concentration in the target materials. The vacuum annealing substantially increased the lateral grain size.

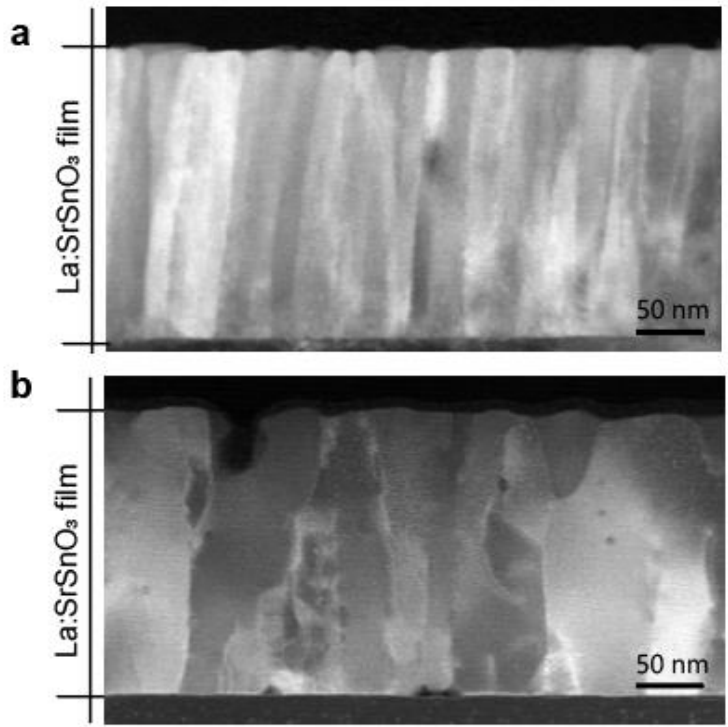

**FIG. 2 | Cross-sectional LAADF-STEM images for the LSSO films.** (a) as grown and (b) annealed LSSO films. The lateral grain growth of annealed LSSO films is clearly visualized.

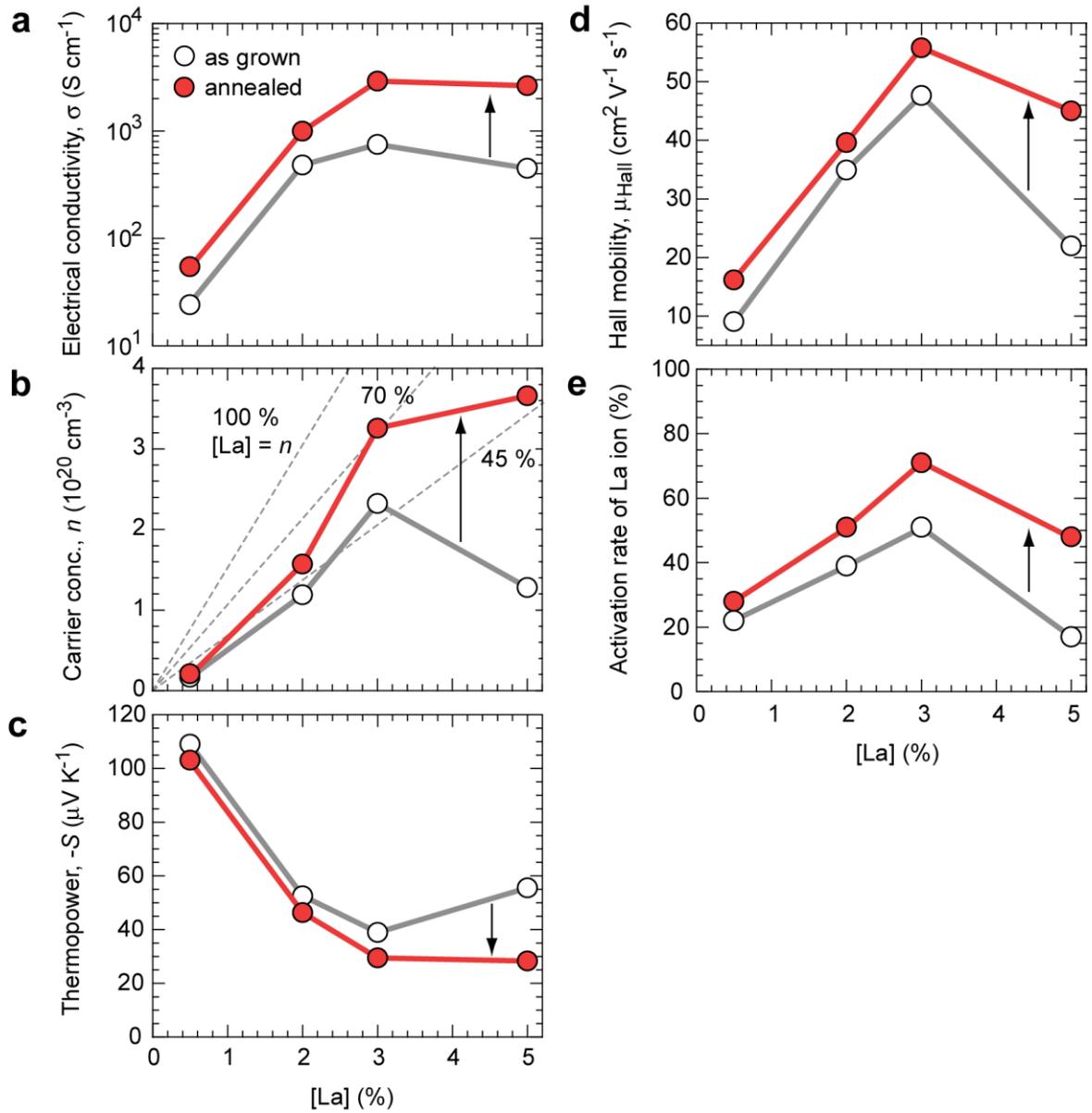

**FIG. 3 | Electron transport properties of the resultant LSSO thin films as a function of La component (%) in the target materials at room temperature.** (a) Electrical conductivity ($\sigma$), enhancement of $\sigma$ for the vacuum annealed LSSO films increased with the La component. (b) Carrier concentration ($n$), $n$ of LSSO film increased after annealing in vacuum. (c) Thermopower ($-S$), the absolute value of thermopower ($S$) decreased after annealing. (d) Hall mobility ($\mu_{Hall}$), the $\mu_{Hall}$ of the annealed LSSO films was largely enhanced at 5% $La^{3+}$. (e) Activation rate of La ion, after vacuum annealing, La ion became more active.

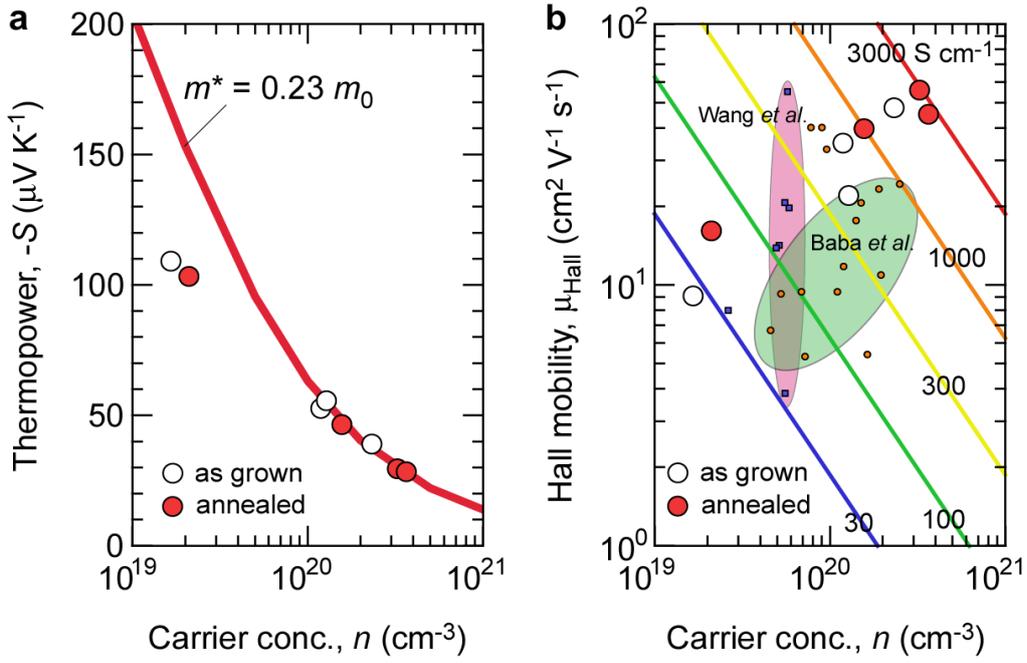

**FIG. 4 | (a) Thermopower (−$S$) and (b) Hall mobility ($\mu_{Hall}$) of resultant LSSO films as a function of carrier concentration ($n$) at room temperature.** The carrier effective mass ($m^*$) values of LSSO is 0.23 $m_0$, clarified by the thermopower measurements. By the thermal annealing in vacuum, we achieved electrical conductivity exceed 3000 S cm$^{-1}$ in LSSO film and the mobility reaches value as high as 55.8 cm$^2$ V$^{-1}$ s$^{-1}$ at a carrier concentration of 3.26 × 10$^{20}$ cm$^{-3}$. The values reported by Wang et al.[20] (blue square) and Baba et al.[10] (orange circle) are also plotted for comparison.

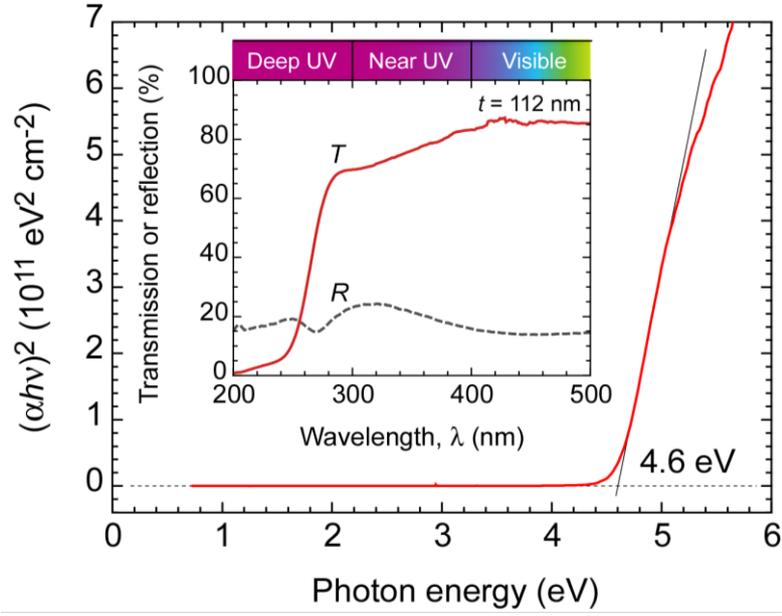

**FIG. 5 | Optical bandgap of the resultant 3% La-doped LSSO thin film.** The $E_g$ was ~4.6 eV, which was obtained from the linear relationship of $(\alpha h\nu)^2$ – photon energy with assuming the direct bandgap. The absorption coefficient $\alpha$ was obtained from the transmission and reflection data shown in the inset.

**Supplementary Material**

**High electrical conducting deep-ultraviolet-transparent oxide semiconductor La-doped SrSnO$_3$ exceeding ~3000 S cm$^{-1}$**

*Mian Wei, Anup V. Sanchela, Bin Feng, Yuichi Ikuhara, Hai Jun Cho\*, and Hiromichi Ohta\**

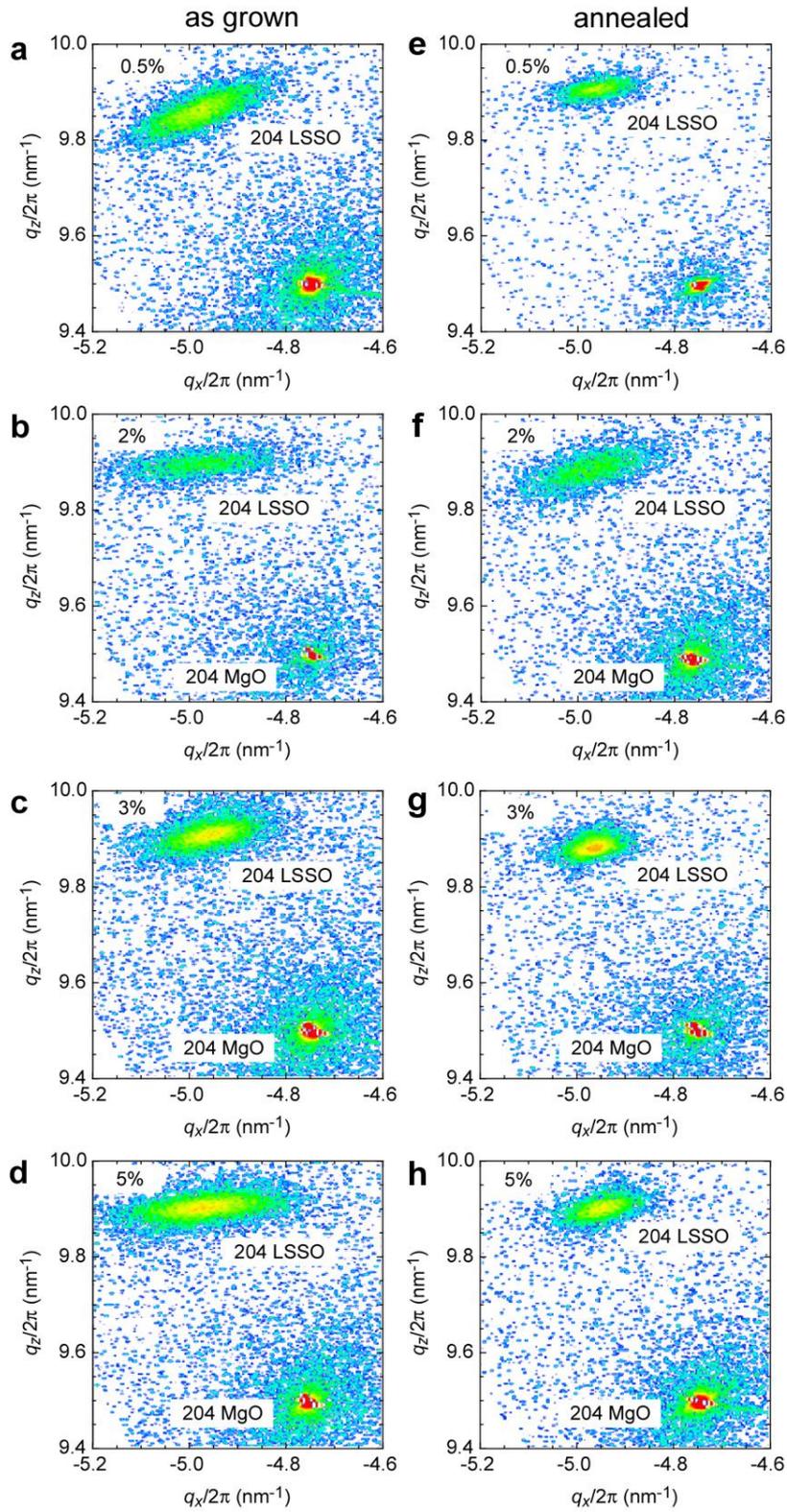

**FIG. S1 | RSM near 204 diffraction spot of the LSSO films.** (a-d) as-deposited film, (e-h) vacuum annealed at 790 °C for 30 min.

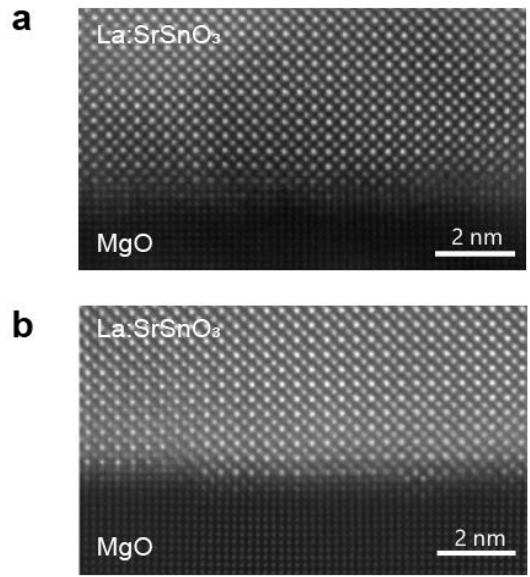

**FIG. S2 | Cross-sectional HAADF-STEM images for the LSSO films.** (a) as grown and (b) annealed LSSO films. Incoherent interfaces with misfit dislocations are seen at the interface between the LSSO film and MgO substrate in both cases. The interfaces are not affected by the vacuum annealing.